\documentclass{Interspeech}

\interspeechcameraready

\title{ABHINAYA -  A System for Speech Emotion Recognition In Naturalistic Conditions Challenge }

\author[affiliation={1}]{Soumya}{Dutta}
\author[affiliation={2}]{Smruthi}{Balaji}
\author[affiliation={1}]{Varada}{R}
\author[affiliation={1}]{Viveka}{Salinamakki}
\author[affiliation={1}]{Sriram}{Ganapathy}
\affiliation{Learning and Extraction of Acoustic Patterns (LEAP) Lab}{Electrical Engineering, Indian Institute of Science, Bangalore}{India}
\affiliation{Shiv Nadar University}{Chennai}{India}
\email{sriramg@iisc.ac.in}

\usepackage{comment}
\usepackage{xcolor}
\usepackage{hyperref}
\usepackage[skins]{tcolorbox}
\usepackage{bbding}
\begin{document}

\maketitle

\begin{abstract}
    Speech emotion recognition (SER) in naturalistic settings remains a challenge due to the intrinsic variability, diverse recording conditions, and class imbalance. As participants in the Interspeech Naturalistic SER Challenge which focused on these complexities, we present ``Abhinaya'', a system integrating speech-based, text-based, and speech-text models. Our approach fine-tunes self-supervised and speech large language models (SLLM) for speech representations, leverages large language models (LLM) for textual context, and employs speech-text modeling with an SLLM to capture nuanced emotional cues. To combat class imbalance, we apply tailored loss functions and generate categorical decisions through majority voting. Despite one model not being fully trained, the Abhinaya system ranked $4$th among $166$ submissions. Upon completion of training, it achieved state-of-the-art performance among published results, demonstrating the effectiveness of our approach for SER in real-world conditions.
    %This paper  details the performance of the components as well as the ablations related to loss function choices. 
\end{abstract}
\section{Introduction}\label{sec:intro}
Speech carries significant para-linguistic information like cues about speaker, language, ambient conditions, emotions, and   physiological/psychological health states of the speaker.  
The efforts to extract   paralinguistic attributes has greatly benefited from evaluation campaigns and global benchmarking, much like other fields of machine learning. For example, speaker and language recognition evaluation challenges held   over two decades~\cite{greenberg2020two, lee23c_interspeech},  speaker diarization challenges~\cite{ryant21_interspeech, baghel2024summary}, COVID-19 diagnostic challenges \cite{muguli2021dicova, sharma2022second}, and speech recognition challenges \cite{nandwana2019voices, watanabe2020chime} have propelled the development of these technologies by fostering community participation. 
% and standardizing evaluation protocols and metrics.
% Beyond the task of speaker identification, the task of recognizing multiple speakers from a single speech signal, also called speaker diarization, has also received considerable interest from the speech research community leading to . The language and speaker identification problems have been combined in the recent DISPLACE challenges~\cite{baghel23_interspeech, kalluri24_interspeech}.

On the task of speech emotion recognition, the conduct of challenges has been relatively sparse. One of the first such attempts for speech emotion recognition (SER), was held in $2009$~\cite{schuller09_interspeech}, which has been extended to multimodal emotion recognition~\cite{schuller2011avec, stappen2020muse}. Most widely used SER datasets are collected in controlled environments, featuring actors expressing scripted emotions~\cite{busso2008iemocap, busso2016msp}. These datasets are typically small, and models trained on them  struggle to generalize to real-world conditions. To address this limitation, the MSP-PODCAST corpus~\cite{lotfian2017building} was introduced as a large-scale, naturalistic dataset. The data formed the basis for the Odyssey $2024$ emotion recognition challenge~\cite{goncalves24_odyssey}. A new version of the MSP-PODCAST dataset is used for the Interspeech $2025$ Emotion Recognition in Naturalistic Conditions Challenge~\cite{Naini_2025}. 

The Naturalistic Conditions Challenge explored the following aspects: 1) Imbalanced training data with balanced test sets across different emotion classes, 2) High variability in emotional expression, featuring more than $2000$ speakers, 3) Availability of text annotations for the training data, enabling the development of multi-modal (speech-text) systems~\cite{dutta2024leveraging}.

This paper describes the details of ``Abhinaya'' (meaning - art of expression in \textit{Sanskrit}) system, an SER model, which was developed as part of this challenge.  The Abhinaya system directly addresses the challenge’s key issues—class imbalance, multimodal modeling, and data variability—through an ensemble of speech-only, text-only, and speech-text joint models. Rather than building these components from scratch, we leverage SSL models (WavLM-Large~\cite{chen2022wavlm}) and speech large language models (SLLMs) (SALMONN~\cite{tang2023salmonn}) as speech feature encoders. Given the potential benefits of multimodal information, we transcribe speech using Whisper-large-v3 automatic speech recognition (ASR) system  \cite{radford2023robust} and process the transcripts with LLMs. Further, we adopt an approach for the multimodal model, where speech and text are concatenated at the input and processed jointly using an SLLM. 
%Our approach uniquely re-purposes an SLLM to model speech-text interactions directly, allowing it to learn rich cross-modal emotional representations within a single framework.
To mitigate class imbalance, we explore various loss functions for fine-tuning LLMs and SLLMs for SER. Overall, the Abhinaya system, based on an ensemble of five models, ranked $4$th out of $166$ entries on the categorical leaderboard\footnote{\url{https://lab-msp.com/MSP-Podcast_Competition/IS2025/}}. 
Notably, one subsystem was not fully fine-tuned at  the official submission deadline. After completion of this fine-tuning, Abhinaya achieved  state-of-the-art (SoTA) performance in post-challenge evaluation.

The key contributions from this work are:
\begin{itemize}
% [leftmargin=*]
    \item
    Exploring Speech Large Language Models (SLLMs) as feature embedding models that can be fine-tuned for the task of emotion recognition from speech or speech-text jointly. 

    \item Investigating specialized loss-functions for large model training in a class-imbalanced setting.

   \item Combining speech-only, text-only and speech-text models with zero-shot/fine-tuned settings using a majority voting rule, achieving SoTA results on the SER classification task in the naturalistic conditions challenge.

\end{itemize}
% This study presents a novel foundation model-based approach to SER, demonstrating the effectiveness of large-scale pre-trained models in capturing both acoustic and semantic cues for robust emotion recognition.

\section{Background}\label{sec:back}
\textbf{Speech foundation models for SER}: Speech foundation models trained in a self-supervised learning fashion such as wav2vec2.0~\cite{baevski2020wav2vec}, HuBERT~\cite{hsu2021hubert} or WavLM~\cite{chen2022wavlm} yield rich speech representations for various tasks, including speech emotion recognition (SER). Both wav2vec2.0 and HuBERT embeddings have been used for SER~\cite{pepino2021emotion,dutta2023hcam, pastor22_iberspeech}. The WavLM-large model was fine-tuned on the MSP-PODCAST dataset~\cite{Naini_2025}, serving as the baseline system for this challenge.
% While emotion data based pre-training (emotion2vec \cite{ma2023emotion2vec}) 

\noindent \textbf{Emotion recognition using LLMs}: A comprehensive evaluation by Zhang et al.~\cite{zhang2024sentiment} highlights the effectiveness of LLMs in text sentiment classification.  However, these models struggle with fine-grained emotion understanding, as shown recently by Dutta et al.~\cite{dutta2025llm}. 
Adapting text-based LLMs for speech tasks with speech-specific encoders 
 has witnessed significant progress in recent years.   Some examples of such speech large language models (SLLMs) are SALMONN~\cite{tang2023salmonn}, Qwen-Audio~\cite{chu2023qwen} and WavLLM~\cite{hu-etal-2024-wavllm}.
 % These models have been shown to be suitable for a number of speech and audio tasks.
  The proposed Abhinaya system uses the SALMONN model for some of the sub-systems.
  
% \noindent \textbf{Loss Functions for imbalanced data}:   A common approach to handling imbalanced datasets is to adjust sample-wise weights based on class distribution. This method, known as weighted cross-entropy (WCE) loss, was used to train the baseline system in this challenge~\cite{Naini_2025}. Another widely used loss function for imbalanced data is the focal loss \cite{lin2017focal}, which dynamically reweights samples based on the model’s prediction confidence. Both these losses were previously explored in the Odyssey 2024 SER Challenge \cite{goncalves24_odyssey}. In our work, we also investigate vector scaling (VS) loss~\cite{kini2021label}, which modifies the model’s predictions using class-wise temperature scaling and bias correction.

\noindent \textbf{Loss Functions for imbalanced data}:   A common strategy to handle imbalanced datasets is using the weighted cross-entropy (WCE) loss, which utilizes sample-wise weights based on the class distribution. This was used in the challenge’s baseline system~\cite{Naini_2025}. Another approach, focal loss~\cite{lin2017focal}, dynamically re-weights samples based on prediction confidence. Both losses were explored in the Odyssey $2024$ SER Challenge~\cite{goncalves24_odyssey}. Additionally, we investigate vector scaling (VS) loss~\cite{kini2021label}, which modifies the model’s predictions using class-wise temperature scaling and bias correction.
\vspace{-0.05in}
\section{Proposed SER System}\label{sec:method}
\begin{figure}
    \centering
    \includegraphics[width=\columnwidth,trim={0cm 5.5cm 0cm 1cm},clip]{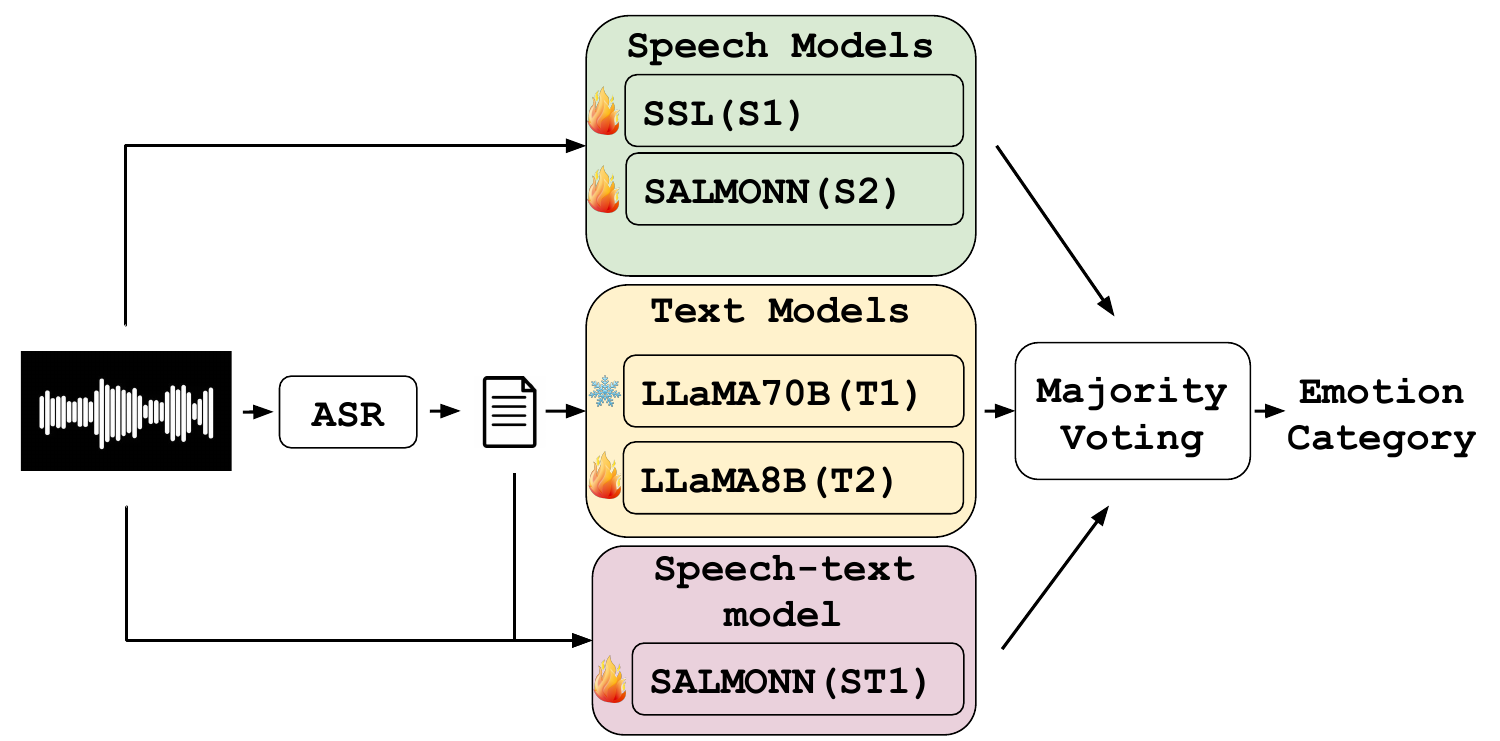}
    \caption{Schematic of the different  components of the Abhinaya SER system. We use three types of models - speech-only (\textbf{S1}, \textbf{S2}), text-only (\textbf{T1}, \textbf{T2}) and speech-text (\textbf{ST1}). Only \textbf{T1} is used in a zero-shot setting. The text used by the models are ASR transcripts   generated by Whisper  \cite{radford2023robust}.
    %\textcolor{purple}{Check if you can increase the text font size by +1 in the figure.}
    %\textcolor{purple}{can you somehow move the Model-V to the bottom of the figure instead of being in the middle.}
    }
    \label{fig:model}
    \vspace{-0.2in}
\end{figure}
% The proposed system consists of three broad categories of models — speech-only models, text-only models and a speech-text joint model.
The schematic of the Abhinaya system\footnote{Code:\url{https://github.com/iiscleap/ABHINAYA}} is shown in Fig.~\ref{fig:model}. 
% We elaborate each of our modeling choices in this section.
\vspace{-0.05in}
\subsection{Speech-only models}
\subsubsection{Wav-LM SSL  (S1)} The speech foundation model used in this sub-system is the   WavLM-large model \cite{chen2022wavlm}. The convolutional feature extractor is frozen, and transformer layers are fine-tuned.  Since WavLM generates frame-level representations, we apply attentive pooling~\cite{okabe18_interspeech} to aggregate these features before emotion classification. The attentive pooling layer outputs a weighted mean of the features along with a weighted standard deviation, where the weights are learned with an attention mechanism.
% \vspace{-0.1in}
\subsubsection{SALMONN SLLM (S2)} 
The SALMONN speech large language model (SLLM) \cite{tang2023salmonn}   
consists of speech encoder - Whisper encoder \cite{radford2023robust} + BEATS encoder  \cite{chen2023beats}, with a text-based LLM (LLaMA \cite{touvron2023llama}). The speech encoder and LLaMA weights remain frozen, while only the  Q-former~\cite{li2023blip} and the low-rank adapters (LoRA) in the LLaMA module are trained for audio and speech tasks  \cite{tang2023salmonn}. 
The SALMONN model is available in two sizes - $7$B and $13$B depending on the size of the back-end text LLM. 

In our setting, we use all the layers of the SALMONN model  except the final text-token layer. 
 The representations from the final LLaMA transformer layer are passed through the attentive statistics pooling layer, and the model is fine-tuned for the emotion classification task with a softmax head. 
 During fine-tuning, we update only the Q-Former and LoRA weights, along with the classification head.
 
 % following which the representations are used for classifying emotions.

% generalize to a number of speech and audio tasks, we fine-tune the SALMONN-$13$B model for the task of emotion recognition. However, since the model is supposed to predict only among $8$ emotion classes, we refrain from treating SALMONN as a text generative model. Instead, we first tokenize the speech signal by Whisper~\cite{radford2023robust} and BEATS~\cite{chen2023beats}. The sequence length of the speech segment is subsequently reduced by a learnable Q-former network~\cite{li2023blip}, following which the LLaMa~\cite{touvron2023llama} model is fine-tuned with the low-rank adaptation (LoRA) strategy~\cite{hu2021lora}. The Q-former network as well as the LoRA weights are initialized from the pre-trained SALMONN-$13$B model. 
\vspace{-0.05in}
\subsection{Text-only models}
In the Naturalistic Conditions Challenge, the reference text transcripts were provided only for the training and validation subsets, while the evaluation data lacked ground-truth transcripts. To ensure consistency, we used ASR transcripts generated by Whisper-Large v3~\cite{radford2023robust} for our model development during training and testing. We explore two choices for text-only models.
% \vspace{-0.1in}
\subsubsection{Zero-shot LLaMA (T1)}\label{sec:llm}
We use the \texttt{LLaMA-3.3-70B-Instruct} model~\cite{dubey2024llama} in zero shot setting for this task.
The model was prompted to classify emotions into predefined categories.
% \vspace{-0.05in}
\subsubsection{Fine-tuned smaller text LLM (T2)} 
Prior studies~\cite{zhang2024sentiment, dutta2025llm} indicate that LLMs often struggle with fine-grained emotion recognition in zero-shot scenario. To address this, we fine-tune the \texttt{LLaMA-3.1-8B} model~\cite{dubey2024llama} using LoRA for emotion recognition.
Similar to the SALMONN fine-tuning setup, we extract last-layer representations from LLaMA, apply attentive statistics pooling, and feed the utterance-level embeddings to an emotion classification head.
\vspace{-0.2in}
\subsection{Speech-text joint model (ST1)}
Since speech large language models (SLLMs) primarily adapt LLMs for speech sequences, we explore their potential for joint speech-text modeling. For this purpose, we use the SALMONN-$7$B model. In addition to the speech sequence, we append the Whisper-generated text transcript before fine-tuning the  SALMONN  model using LoRA. As in speech/text only fine-tuning, we extract final-layer representations, apply attentive statistics pooling, and  train the classification head.
\vspace{-0.05in}
\subsection{Loss functions}
The dataset is highly imbalanced with the frequency of the majority class (``neutral'') almost $26$ times that of the minority class (``fear'') in the training data.  To address this issue, we explore various loss functions:
% \vspace{-0.1in}
% \vspace{-0.05in}
\subsubsection{Weighted cross entropy  (WCE)} \label{sec:wce} 
% In this loss, the contribution of a sample  to the overall loss is scaled based on the frequency of the class it belongs to. 
Denoting the total number of data points by $N$, the number of classes by $C$, the weight assigned for a particular class, $c \in [1,2,\dots,C]$, is given by $w^c = \frac{N}{N_c \times C}$, where there are $N_c$ samples belonging to class $c$ in the training data. The cross-entropy loss $l_{i}$ for sample $i$ is given by:
\begin{equation}\label{eq:wce}
    l_{i} = -\sum_{c=1}^{C}t_{ic}w^c\log(p_{ic})~~~;~~~ {\mathbb E} = \frac{1}{N}\sum_{i=1}^N l_{i}
\end{equation}
where $p_{ic}$ is the prediction probability of class $c$ for sample $i$. $t_{ic}$ is set to $1$ for the true class $c$ of sample $i$, else it is kept at $0$. The  loss $\mathbb E$ is defined as the weighted cross-entropy loss (WCE). 
% \vspace{-0.1in}
% \vspace{-0.2in}
\subsubsection{Weighted FoCal Loss (WFL)} \label{sec:wfl}
To further emphasize hard-to-classify samples, we explore the focal loss~\cite{lin2017focal}, which adjusts the weight based on the model’s confidence. 
% The probabilities, $p_y^c$, are used to dynamically adjust sample weights. 
The weighted focal loss (WFL) is defined as:
\begin{equation}\label{eq:weightedfocal}
    l_{i} = -\sum_{c=1}^{C}t_{ic}w^c(1-p_{ic})^{\gamma}\log(p_{ic}), 
\end{equation}
and the total loss $\mathbb E$ is the same as in Eq.~\ref{eq:wce}. Here, $\gamma \geq 0$ is a hyper-parameter.
% \vspace{-0.1in}
% \vspace{-0.05in}
\subsubsection{Vector scaling (VS) loss} 
%Re-weighting samples based on the class frequencies can lead to over-fitting on the minority samples in the training data~\cite{ye2020identifying}. In order to compensate for this, Ye et al.~\cite{ye2020identifying} propose utilizing a class-dependent temperature to modify the model predictions. This temperature term increases with the frequency of the class, thereby reducing the prediction value for the minority classes relative to the majority class. Further, Kini et al.~\cite{kini2021label} show that adding a class dependent factor to the temperature-modified prediction value further alleviates the learning issues encountered by the model in imbalanced datasets. 

The vector scaling (VS) loss adjusts predictions using class-dependent temperature scaling and bias correction. It modifies the pre-softmax model prediction, $z_{ic}$, as:
\begin{equation}\label{eq:adjust}
    \hat{z}_{ic} = (\frac{N_{c}}{N_{max}})^{\gamma}z_{ic} + \tau \log(\frac{N_c}{N})
\end{equation}
where $\gamma \geq 0$ and $\tau \geq 0$ are hyper-parameters, while $N_{max}=\max_{c\in[1,2,\dots,C]} N_c$. 
The softmax $\hat{p}_{ic} = softmax(\hat{z}_{ic})$ is used in the weighted loss:
\begin{equation}\label{eq:vs}
    l_{i} = -\sum_{c=1}^{C}t_{ic}w^c\log(\hat{p}_{ic})
\end{equation}
where $w^c$ and $t_{ic}$ are defined as before. The total loss ($\mathbb E$) is the same as in Eq.~\ref{eq:wce}.
\vspace{-0.05in}
\section{Experiments}\label{sec:results}
\subsection{Dataset}
The MSP-PODCAST~\cite{lotfian2017building} dataset serves as the basis for this challenge. The training, validation and the test splits contain $84260$, $31961$ and $3200$ speech files respectively. The dataset includes annotations for $8$ primary emotion categories: ``happy'', ``angry'', ``sad'', ``neutral'', ``surprise'', ``fear'', ``contempt'' and ``disgust''. Additionally, two other labels—``other'' (denoting emotions outside the predefined categories) and ``X'' (cases with no annotator consensus)—are present but excluded from emotion modeling. The training and the validation distribution for the $8$ categories are reported in Naini et al.~\cite{Naini_2025}. For the model selection, we construct a balanced validation set, using the same number of utterances ($326$) for each of the $8$ emotion classes. 
The test data labels in the challenge were not provided and the performance of the models was only available through the leaderboard. 
% Since ``fear'' has only $326$ samples in the validation data, an equal number of utterances ($326$) is randomly selected from each of the other $7$ emotion classes. 
% This results in a balanced validation split containing $2608$ samples.
\vspace{-0.05in}
\subsection{Implementation Details}
Except for T1, which operates in a zero-shot manner, all models are fine-tuned using the AdamW optimizer~\cite{loshchilov2017decoupled} with a learning rate of $1e\text{-}5$. For models utilizing LoRA (S2, T2, and ST1), the LoRA parameters~\cite{hu2021lora} are set as $r=8$ and $\alpha=32$, with a dropout rate of $0.1$. Training configurations vary across models due to the diverse memory requirements:
\begin{itemize}
    \item For S1 and T2, the batch size is set to $8$ with gradient accumulation over $4$ steps.
    \item For SALMONN-based models (S2 and ST1), the batch size is set to $4$ with gradient accumulation over $8$ steps.
\end{itemize}
 In speech-only models (S1 and S2) and the speech-text model (ST1), audio samples are processed with a maximum duration of $10$ seconds. All four models are trained for $20$ epochs, with the checkpoint yielding the highest validation macro-F1 score selected for evaluation. Regarding loss functions, models trained with WFL (Eq.~\ref{eq:weightedfocal}) use $\gamma=2$, while those trained with VS loss (Eq.~\ref{eq:adjust}) have hyper-parameters  $\gamma=0.3$ and $\tau=1$. \\
Once the outputs from the five models are available, majority voting is applied across the five classifiers, with S2 serving as the tiebreaker due to its superior validation performance.
\begin{table}[t]
\centering
\caption{The performance (in $\%$) of the different models on the balanced validation and test data. 
The performance of certain models on the test data are not reported as they were not submitted to the leaderboard. 
%WFL stands for weighted focal loss, while WCE stands for weighted cross-entropy loss. 
$*$ indicates the test score evaluated by the organizers after the official challenge deadline.}\label{tab:results}
\renewcommand{\arraystretch}{1.2}
 \resizebox{\columnwidth}{!}{%
\begin{tabular}{@{}l|l|l|l|l@{}}
\toprule
Models & Modality & Loss & \begin{tabular}[c]{@{}l@{}}Val. F1 \\ (macro)\end{tabular} & \begin{tabular}[c]{@{}l@{}}Test F1\\ (macro)\end{tabular} \\ \midrule
Baseline~\cite{Naini_2025} & Speech & WCE & - & $32.93$ \\ \midrule
\textbf{S1} (WavLM-large-317M) & Speech & WFL & $34.43$ & $33$ \\
\textbf{S2} (SALMONN-13B) & Speech & WFL & $37.68$ & $35.34$ \\
\textbf{T1} (LLaMA-3.3-70B)& Text & - & $32.78$ & - \\
\textbf{T2} (LLaMA-3.1-8B) & Text & VS & $33.68$ & - \\
\textbf{ST1} (SALMONN-7B) & Speech-text & VS & $35.43$ & - \\ \midrule \midrule
ABHINAYA (4th place) & Speech-text & - & $\mathbf{43.01}$ & $41.81$ \\
ABHINAYA$^{*}$ (SoTA)  & Speech-text & - & $42.31$ & $\mathbf{44.02}$ \\ \bottomrule

\end{tabular}}
\vspace{-0.2in}
\end{table}
\begin{figure}[t]
    \centering
    \includegraphics[width=\columnwidth,trim={5cm 7.5cm 6cm 3cm},clip]{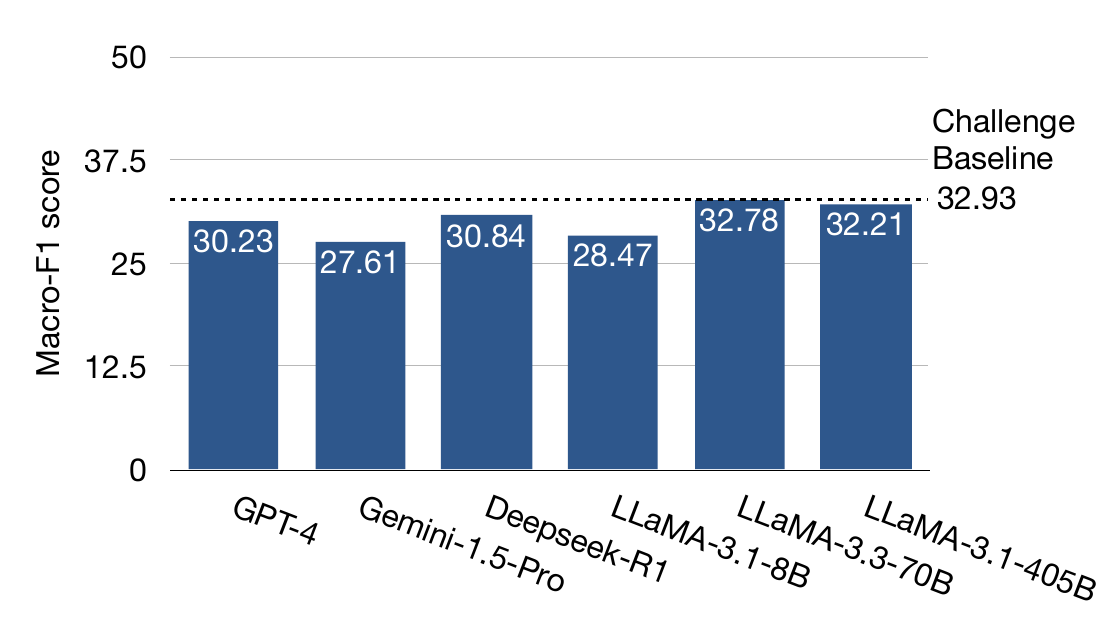}
    \caption{Validation macro F1-score (in $\%$) for different LLMs~\cite{dubey2024llama, guo2025deepseek, team2023gemini, achiam2023gpt} evaluated in zero-shot setting using the ASR transcripts. The LLaMA models considered are the Instruct versions. The baseline performance is also shown.}
    \label{fig:llms}
    \vspace{-0.25in}
\end{figure}
% \footnotetext[3]{\url{https://openai.com/index/gpt-4/}}
\vspace{-0.05in}
\subsection{Performance of Abhinaya}\label{sec:results_final}
Table~\ref{tab:results} presents the validation and test set results. The key findings are summarized below:
\begin{itemize}
    \item \textbf{Fine-tuned SALMONN-13B (S2) achieves the best individual system performance}:  Note that, its zero-shot validation performance is only $18.63\%$, hence fine-tuning significantly improves the performance.
    \item \textbf{Fine-tuning a small model versus zero-shot evaluation of a large model}: The fine-tuned LLaMA-3.1-8B (T2) model outperforms the zero-shot LLaMA-3.3-70B model (T1).
    \item \textbf{Ensembling improves performance}: A majority vote ensemble of the five models achieves a $24.56\%$ relative gain over the best individual model (S2) on the test set.
    \item \textbf{Significant improvement over the challenge baseline}: Our system achieves a $33.68\%$ relative improvement over the baseline, demonstrating its effectiveness for naturalistic SER.
    \item \textbf{Post-challenge evaluation confirms state-of-the-art performance}: At the challenge submission deadline, the ST1 fine-tuning had  completed only $3$ epochs. We continued this model fine-tuning till completion ($20$ epochs) and the organizers performed a post-challenge evaluation of the Abhinaya system with this update.  The final system performance of $44.02\%$ surpasses the best published result in the challenge, thereby providing the SoTA results on this task.
\end{itemize}
%Among all models, the fine-tuned SALMONN-$13$B model (S2) achieves the best performance. Notably, its fine-tuning plays a crucial role, as its zero-shot validation performance is only $18.63\%$.
% We also observe that the fine-tuned \texttt{LLaMA-3.1-8B} model (T2) outperforms the bigger \texttt{LLaMA-3.3-70B-Instruct} model (T1), highlighting the effectiveness of fine-tuning over relying on larger-scale zero-shot models. It is to be mentioned that the comparison between models S1 and ST1 is not fair as we fine-tune the SALMONN-$13$B model in the former in comparison to the SALMONN-$7$B model in the latter. 
 %A majority vote among these $5$ models improves over the best performing single model (model S2) by $24.56\%$ (relative) on the test data, which indicates the advantage of taking an ensemble of multiple models. Further, our system achieves a relative improvement of $33.68\%$ over the baseline model indicating the applicability of our proposed system for the task of naturalistic SER. Note that due to the challenge deadline, we had to use a partially trained model ST1 for our ensemble, and hence the latest leaderboard entry is $41.81\%$. However, post the conclusion of the challenge, the organizers evaluated our model ensemble and confirmed our model to be the state-of-the-art in this challenge (improving upon the nearest team by $2\%$ relative).
% \subsection{Discussion}

\vspace{-0.05in}
\subsection{Evaluation of different LLMs for text-only models}\label{sec:evalllm}
For the T1 zero-shot setting, we evaluate multiple large language models (LLMs) for the task of emotion recognition using a same prompt structure.  
% We note that \texttt{LLaMA-3.3-70B-Instruct} outperforms all the others. 
The comparative performance on the validation set is presented in Fig.\ref{fig:llms}. The key observations are:
\begin{itemize}
    \item LLaMA-3.3-70B-Instruct achieves the best performance, outperforming all other LLMs considered here.
    \item While the zero-shot performance of LLaMA-3.1-8B-Instruct model is only $28.47\%$, the fine-tuning with VS loss improves the performance to $33.68\%$ (Table~\ref{tab:results}). 
    \item The other ``larger'' LLMs did not perform as well on the emotion recognition task with ASR transcripts.
\end{itemize}
\begin{table}[t!]
\centering
\caption{The performance of the models with different loss functions. All results are macro F1-scores ($\%)$ on the balanced validation data. \textbf{Bold} indicates best loss function for each model.}\label{tab:loss}
\renewcommand{\arraystretch}{1.2}
 \resizebox{0.5\columnwidth}{!}{%
\begin{tabular}{@{}c|c|c|c@{}}
\toprule
Models & WCE & WFL & VS \\ \midrule
S1 & $33.07$ & $\mathbf{34.43}$ & $32.12$ \\
S2 & $36.34$ & $\mathbf{37.68}$ &  $33.17$\\
T2 & $29.79$ & $30.12$ & $\mathbf{33.68}$ \\ 
ST1 & $33.92$ & $34.73$ & $\mathbf{35.43}$ \\ \bottomrule
\end{tabular}}
\vspace{-0.2in}
\end{table}
\vspace{-0.05in}
\subsection{Choice of loss functions}
The impact of different loss functions is analyzed in Table~\ref{tab:loss}. The key observations are:
\begin{itemize}
    \item Speech-only models (S1, S2) benefit most from weighted focal loss, suggesting its effectiveness in handling class imbalance for SER. In contrast, VS loss is effective for the text-based (T2) and speech-text (ST1) models.
    \item The stronger zero-shot performance of T2 ($28.47\%$ versus $18.63\%$ for S2) suggests better initial separation between emotion classes. We hypothesize that this initial separation aids VS loss, which adjusts logits class-wise, whereas speech models with less discriminative initial representations benefit more from sample-wise loss weighting.
\end{itemize}
\vspace{-0.05in}
\subsection{Speech-text versus speech-only SLLMs}
To assess the impact of joint speech-text fine-tuning, we fine-tune a SALMONN-$7$B model using speech inputs. This setup mirrors model S2 but replaces the SALMONN-$13$B model with the smaller $7$B variant for a direct comparison with ST1 model, which incorporates both speech and text. The speech-only model achieves $33.87\%$ validation macro-F1, while ST1 achieves $35.43\%$, demonstrating the benefits of multimodal fine-tuning. However, speech-text models incur higher computational costs due to the longer input sequences.

%%%%%%%%%%%%%%%%%%%%%%%%%%%%%%%%%
%Am not sure if this is adding anything to the paper. So removed it by the heatmap section
%%%%%%%%%%%%%%%%%%%%%%%%%%%%%%%%%

\begin{table}[t]
\centering
\caption{Class-wise macro F1-scores ($\%$) for the balanced validation data for Abhinaya and the different model components. \textbf{Bold} indicates best model for each emotion category.}\label{tab:class}
\renewcommand{\arraystretch}{1.2}
 \resizebox{\columnwidth}{!}{%
\begin{tabular}{@{}l|c|c|c|c|c||c@{}}
\toprule
\begin{tabular}[c]{@{}l@{}}Emotion\\ Classes\end{tabular} & S1 & S2 & T1 & T2 & ST1 & ABHINAYA \\ \midrule
Angry & $49.74$ & $50.60$ & $32.48$ & $37.13$ & $50.73$ & $\mathbf{51.49}$ \\
Contempt & $20.02$ & $13.74$ & $22.37$ & $22.56$ & $\mathbf{25.62}$ & $22.87$ \\
Disgust & $18.91$ & $31.53$ & $33.33$ & $\mathbf{37.31}$ & $32.08$ & $36.13$ \\
Fear & $12.22$ & $16.79$ & $\mathbf{29.22}$ & $27.74$ & $29.09$ & $26.41$ \\
Happy & $55.93$ & $\mathbf{62.94}$ & $40.00$ & $43.33$ & $50.88$ & $62.83$ \\
Neutral & $33.13$ & $37.06$ & $31.24$ & $20.45$ & $16.37$ & $\mathbf{40.29}$ \\
Sad & $50.60$ & $51.47$ & $40.69$ & $47.47$ & $46.86$ & $\mathbf{55.62}$ \\
Surprise & $34.94$ & $37.31$ & $32.93$ & $33.48$ & $31.83$ & $\mathbf{42.81}$ \\ \bottomrule
\end{tabular}}
\vspace{-0.1in}
\end{table}

% \begin{figure}
%     \centering
%     \includegraphics[width=\columnwidth,trim={3.5cm 2cm 5cm 3cm},clip]{figures/heatmap.pdf}
%     \caption{Confusion matrix for the balanced validation data for the Abhinaya$^*$ system. The labels are, A:angry, C:contempt, D:disgust, F:fear, H:happy, N:neutral, S:sad and U:surprise.}
%     \label{fig:heatmap}
%     \vspace{-0.2in}
% \end{figure}
\vspace{-0.05in}
\subsection{Class-wise performance analysis}
Table~\ref{tab:class} presents macro-F1 scores per class for Abhinaya and its components. The speech models (S1, S2) perform poorly on the least frequent classes~\cite{Naini_2025}: ``fear'', ``contempt'', ``disgust''. In contrast, the fine-tuned text model (T2) and the speech-text model (ST1) show improved performance on these classes.  Interestingly, despite its weaker overall performance (Table~\ref{tab:results}), the zero-shot text model (T1) outperforms the speech-based models on the three rare classes and even outperforms the ensemble for the least frequent class, ``fear''. The speech-text model (ST1) generally performs between the speech and text models across classes (except ``neutral''), while the ensemble achieves the best scores in four out of eight emotion categories.

%The confusion matrix on  the balanced validation set for the proposed Abhinaya$^*$ system (Fig.~\ref{fig:heatmap}) reveals the key classification challenges in this task. 
%In-spite of the efforts undertaken in the system design to improve the training of classes which are under-represented in the training data, 
%the model exhibits significant difficulty in classifying emotion classes like ``Contempt'', ``Disgust'' and ``Fear''. 
%This may be attributed to the lack of diversity of training speakers (due to the limited number of samples) as well as the labeling difficulties in  fine-grained emotion classes. 
%More efforts on data augmentation, mix-up training and synthetic data generation may likely suppress these effects, and these form part of our future exploration. 
% differentiate angry (A), disgust (D), and contempt (C), likely due to their overlapping acoustic properties. Additionally, fear (F), the least frequent class, is often misclassified as neutral (N), the most dominant category, highlighting the impact of dataset imbalance.
% Despite these challenges, the model performs well on the remaining five emotion classes, demonstrating its effectiveness in capturing a broad range of emotions while still facing fine-grained classification difficulties.

\begin{table}[t]
\centering
\caption{Results (in $\%$) with majority voting from different combinations of models from Abhinaya.}\label{tab:majority}
\resizebox{0.8\columnwidth}{!}{%
\begin{tabular}{@{}l|c|c|c|c|c||c|c@{}}
\toprule
\begin{tabular}[c]{@{}l@{}}Model\\ ensemble\end{tabular} & S1 & S2 & T1 & T2 & ST1 & \begin{tabular}[c]{@{}l@{}}Val. F1\\ (macro)\end{tabular} & \begin{tabular}[c]{@{}l@{}}Test F1\\ (macro)\end{tabular} \\ \midrule
Comb. I & \XSolid & \Checkmark  & \XSolid & \Checkmark & \Checkmark & $40.36$ & $39.74$ \\ \midrule
Comb. II & \Checkmark & \XSolid  & \XSolid & \Checkmark & \Checkmark & $39.07$ & $-$ \\ \midrule
Comb. III & \Checkmark & \Checkmark & \XSolid & \Checkmark & \Checkmark  & $40.99$ & $-$ \\ \midrule 
Comb. IV & \Checkmark & \Checkmark & \Checkmark & \Checkmark & \XSolid  & $40.48$ & $-$ \\ \midrule \midrule
ABHINAYA & \Checkmark & \Checkmark  & \Checkmark  & \Checkmark  & \Checkmark  & $42.31$ & $44.02$ \\ \bottomrule
\end{tabular}}
\vspace{-0.2in}
\end{table}
\vspace{-0.05in}
\subsection{Majority voting with other combinations}
We evaluate model combinations using majority voting (Table~\ref{tab:majority}). The best-performing trio—S2 (speech), T2 (text), and ST1 (speech-text)—achieves $40.36\%$ validation macro-F1 and $39.74\%$ on the test set (Comb. I). Adding S1 improves validation performance by $0.63\%$ absolute (Comb. III), while ensembling all five models boosts it by nearly $1.5\%$ absolute (ABHINAYA). Since T1, T2, and ST1 perform better on the minority classes, ensembling all models enhances performance on these underrepresented categories compared to ensembling only S1, S2, T2, and ST1. The impact of ST1 is evident from the nearly $2\%$ absolute drop in Comb. IV when it is removed.  We also evaluate majority voting with the three models having the fewest parameters (S1, T2, and ST1) (Comb. II), using ST1 for tie-breaking. The validation F1 score drops to $39.07\%$, highlighting the need for larger models for the SER task.

%Note that for all the combinations in Table~\ref{tab:majority}, predictions from model S2 are considered in case of no absolute majority.
\vspace{-0.05in}
\section{Conclusion}
In this paper, we present Abhinaya, a system for speech emotion recognition (SER) developed as part of the Interspeech Naturalistic Speech Emotion 2025 challenge. Our approach combines five models leveraging large language models (LLMs) and speech large language models (SLLMs)—two speech-based, two text-based, and one multimodal model. To address class imbalance, we explore various loss functions, demonstrating their impact on model performance. Our model ensemble achieves the best published results on this task. Experimental analysis highlights   the importance of the different components and loss functions suitable for fine-tuning each model.
\bibliographystyle{IEEEtran}
\bibliography{refs}

\end{document}